\documentclass[aps,prl,twocolumn, showpacs,
  superscriptaddress]{revtex4}

\usepackage{graphicx} \usepackage{color} \usepackage{natbib}

\begin{document}

   \title{Measuring neutron-star properties via gravitational waves from neutron-star mergers}

\author{A.~Bauswein} \affiliation{Max-Planck-Institut f\"ur
  Astrophysik, Karl-Schwarzschild-Str.~1, D-85748 Garching, Germany}
\author{H.-T.~Janka} \affiliation{Max-Planck-Institut f\"ur
  Astrophysik, Karl-Schwarzschild-Str.~1, D-85748 Garching, Germany}

	\date{\today} 

  \begin{abstract}
We demonstrate by a large set of merger simulations for
symmetric binary neutron stars (NSs) that there is a tight correlation
between the frequency peak of the postmerger gravitational-wave (GW) emission and the
physical properties of the nuclear equation of state (EoS), e.g. expressed
by the radius of the maximum-mass Tolman-Oppenheimer-Volkhoff
configuration. Therefore, a single measurement of the peak frequency
of the postmerger GW signal will constrain the NS EoS significantly.
For optimistic merger-rate estimates a corresponding detection with Advanced LIGO is expected to happen within
an operation time of roughly a year.

   \end{abstract}


   \maketitle

The properties of high-density matter as in the cores of NSs, in particular the EoS, are still incompletely known, because the physical conditions are not directly accessible by experiments. Theoretical models for supernuclear matter are ambiguous and suffer from uncertainties of nuclear data required as input for these calculations~\cite{2007ASSL..326.....H}.

NS properties are intimately linked to the adopted EoS because the latter determines the stellar structure by the Tolman-Oppenheimer-Volkhoff (TOV) equations \cite{2007ASSL..326.....H,2007PhR...442..109L}. Hence, constraints on the NS EoS can be deduced from astrophysical observations (e.g.~\cite{2010Natur.467.1081D}), as alternatives to nuclear models \cite{2010PhRvL.105p1102H} and laboratory experiments~\cite{2007PhR...442..109L}.

NS mergers may also yield information about the nuclear EoS, because the dynamics of the coalescence depend sensitively on the behavior of high-density matter (see~\cite{2010CQGra..27k4002D,2009CQGra..26k4004F} for reviews). Consequently, the EoS leaves an imprint on the GW signal of NS mergers. However, the systematic dependences of the inverse problem, i.e. which EoS (or NS) properties can be derived from a particular GW detection, are still not completely explored (see~\cite{1994PhRvD..50.6247Z,2002PhRvL..89w1102F,2005PhRvL..94t1101S,2005PhRvD..71h4021S,2006PhRvD..73f4027S,2007PhRvL..99l1102O,2009PhRvD..79l4033R,2010CQGra..27k4002D,2010PhRvL.105z1101B,2011PhRvD..83d4014G,2011PhRvD..83l4008H} and refs. therein). In this letter we report on a tight correlation between NS parameters and thus EoS characteristics and the dominant frequency of the postmerger GW emission revealed by a systematic study with 18 microphysical EoSs. Our survey is in particular important because the second-generation interferometric GW detectors of Adv. LIGO \cite{2010CQGra..27h4006H} and Adv. Virgo \cite{Acernese:2006bj} go into operation within the next years. NS binaries are considered a major target of these instruments with an estimated detection rate of 0.4 to 400/yr \cite{2010CQGra..27q3001A}.

Our simulations are performed with a 3-D relativistic smoothed particle hydrodynamics (SPH) code, which solves the Einstein field equations assuming conformal flatness and employing a GW backreaction scheme within a post-Newtonian framework~\footnote{Testing the GW backreaction scheme by comparing the inspiral times from a certain orbit to maximum compression with results of~\cite{2005PhRvD..71h4021S,2006PhRvD..73f4027S,2011PhRvD..83l4008H}, we find agreement to better than $\sim 25~\%$.}~\cite{2007A&A...467..395O,PhysRevD.81.024012}. The implementation allows the usage of tabulated microphysical EoSs including thermal effects, or arbitrary barotropic EoSs (e.g. zero-temperature EoSs for equilibrium to weak interactions, the so-called $\beta$-equilibrium). The latter are supplemented by an ideal-gas component with an ideal-gas index $\Gamma_{\mathrm{th}}=2$ to mimic thermal effects~\cite{2010PhRvD..82h4043B}.

The calculations start from quasi-equilibrium orbits about two revolutions before the merging of the NSs, which are assumed to be initially cold and in neutrino-less $\beta$-equilibrium. Because tidally locked binaries are unlikely to occur~\cite{1992ApJ...400..175B}, the stars are set up as nonrotating, which is a valid approximation even for millisecond NSs, whose rotation is still slow compared to the orbital period. If not noted otherwise the NSs are modeled by about 340,000 SPH particles.

\begin{figure}
\includegraphics[width=8.9cm]{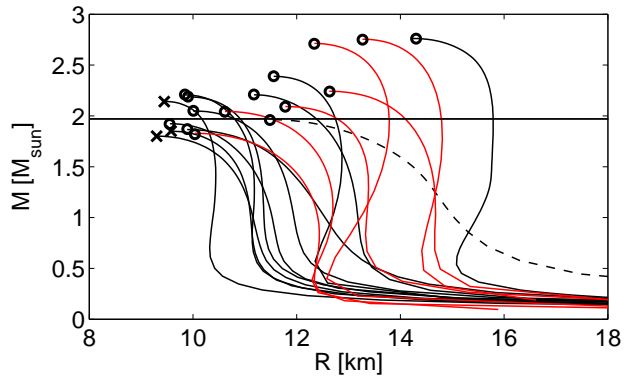}
\caption{\label{fig:tov} NS $M$-$R$ relations for all considered EoSs. Red curves correspond to EoSs that include thermal effects consistently, black lines indicate EoSs supplemented with a thermal ideal gas. The horizontal line corresponds to the 1.97~$M_{\odot}$ NS~\cite{2010Natur.467.1081D}.}
\end{figure}

In total we employ 18 different microphysical EoSs (see Tab.~\ref{tab:models} for the nomenclature and references). Seven of these EoSs include thermal effects consistently. The remaining ones describe nuclear matter at zero temperature and are labeled with ``$+\Gamma_{\mathrm{th}}$'' in Tab.~\ref{tab:models}. The mass-radius ($M$-$R$) relations, the maximum masses $M_{\mathrm{max}}$ of nonrotating NSs and the corresponding (minimum) radii, denoted as $R_{\mathrm{max}}$, for all used EoSs are shown in Fig.~\ref{fig:tov}. The maximum-mass configurations (Tab.~\ref{tab:models}) are marked by symbols. The scatter in Fig.~\ref{fig:tov} illustrates the diversity of the microphysical models underlying our study.

\begin{table}
\caption{\label{tab:models} Used EoSs. $M_{\mathrm{max}}$ and $R_{\mathrm{max}}$ are mass and radius of the maximum-mass TOV configuration, $f_{\mathrm{peak}}$ is the peak frequency of the postmerger GW emission with the FWHM (a cross indicates prompt collapse of the remnant). $f \tilde{h}_{z}(f_{\mathrm{peak}})$ is the effective peak amplitude of the GW signal at a polar distance of 20~Mpc. The tables of the first five and next seven EoSs are taken from \cite{refLorene} and \cite{refRNS}, respectively.}
 \begin{ruledtabular}
 \begin{tabular}{|l|l|l|l|l|}
EoS with    & $M_{\mathrm{max}}$ & $R_{\mathrm{max}}$ & $f_{\mathrm{peak}}$, FWHM  & $f \tilde{h}_{z}(f_{\mathrm{peak}})$ \\
references  & $[M_{\odot}]$      & [km]               & [kHz]                & $[10^{-21}]$               \\
\hline
Sly4  \cite{2001AA...380..151D}  $+\Gamma_{\mathrm{th}}$  & 2.05          &  10.01       & 3.32, 0.20 & 2.33 \\
APR   \cite{1998PhRvC..58.1804A} $+\Gamma_{\mathrm{th}}$  & 2.19          &  9.90        & 3.46, 0.18 & 2.45 \\
FPS   \cite{1981NuPhA.361..502F} $+\Gamma_{\mathrm{th}}$  & 1.80          &  9.30        & x          & x    \\
BBB2  \cite{1997AA...328..274B}  $+\Gamma_{\mathrm{th}}$  & 1.92          &  9.55        & 3.73, 0.22 & 1.33 \\
Glendnh3 \cite{1985ApJ...293..470G}$+\Gamma_{\mathrm{th}}$& 1.96          &  11.48       & 2.33, 0.13 & 1.27 \\
eosAU \cite{1988PhRvC..38.1010W} $+\Gamma_{\mathrm{th}}$  & 2.14          &  9.45        & x          & x    \\
eosC  \cite{1974NuPhA.230....1B} $+\Gamma_{\mathrm{th}}$  & 1.87          &  9.89        & 3.33, 0.22 & 1.27 \\
eosL  \cite{1975PhLB...59...15P} $+\Gamma_{\mathrm{th}}$  & 2.76          &  14.30       & 1.84, 0.10 & 1.38 \\
eosO  \cite{1975PhRvD..12.3043B} $+\Gamma_{\mathrm{th}}$  & 2.39          &  11.56       & 2.66, 0.11 & 2.30 \\
eosUU \cite{1988PhRvC..38.1010W} $+\Gamma_{\mathrm{th}}$  & 2.21          &  9.84        & 3.50, 0.17 & 2.64 \\
eosWS \cite{1988PhRvC..38.1010W} $+\Gamma_{\mathrm{th}}$  & 1.85          &  9.58        & x          & x    \\
SKA   \cite{1991NuPhA.535..331L} $+\Gamma_{\mathrm{th}}$  & 2.21          &  11.17       & 2.64, 0.13 & 1.96 \\
Shen  \cite{1998NuPhA.637..435S}                          & 2.24          &  12.63       & 2.19, 0.15 & 1.43 \\
LS180 \cite{1991NuPhA.535..331L}                          & 1.83          &  10.04       & 3.26, 0.25 & 1.19 \\
LS220 \cite{1991NuPhA.535..331L}                          & 2.04          &  10.61       & 2.89, 0.21 & 1.63 \\
LS375 \cite{1991NuPhA.535..331L}                          & 2.71          &  12.34       & 2.40, 0.13 & 1.82 \\
GS1   \cite{2011PhRvC..83c5802S}                          & 2.75          &  13.27       & 2.10, 0.12 & 1.46 \\
GS2   \cite{2011PhRvC..83f5808S}                          & 2.09          &  11.78       & 2.53, 0.12 & 2.15 \\
 \end{tabular}
 \end{ruledtabular} 
\end{table}

We consider EoSs with $M_{\mathrm{max}}$ in the range of
1.80~$M_{\odot}$ to 2.76~$M_{\odot}$ and $R_{\mathrm{max}}$ from
9.30~km to 14.30~km without any special selection procedure except that we require $M_{\mathrm{max}}\ge 1.8~M_{\odot}$. The lower limit of 1.8~$M_{\odot}$ is motivated by the detection of a pulsar with a mass of $(1.97 \pm 0.04)~M_{\odot}$ \cite{2010Natur.467.1081D}. Although this observation rules out some EoSs of our sample, we do not disregard these models, because at lower densities (as present in 1.35~$M_{\odot}$ NSs and in the merger remnant where strong rotational and thermal effects come into play) these EoSs may still provide a viable description of nuclear matter. Furthermore, the inclusion of these EoSs demonstrates the validity of the relations between merger and EoS properties discussed below over a wider parameter range.

For each EoS listed in Tab.~\ref{tab:models} we simulate the merger of two stars with 1.35~$M_{\odot}$. This setup is chosen because pulsar observations and population synthesis studies suggest these systems to be most abundant~\cite{1999ApJ...512..288T}. After energy and angular momentum losses by GWs have driven the inspiral of the NSs for several 100 Myrs, there are two different outcomes of the coalescence. Either the two stars directly form a black hole (BH) shortly after they fuse (``prompt collapse''), or the merging leads to the formation of a differentially rotating object (DRO) that is stabilized against the gravitational collapse by rotation and thermal pressure contributions. Continuous loss of angular momentum by GWs and redistribution to the outer merger remnant will finally lead to a ``delayed collapse'' on timescales of typically several 10--100~ms depending on the mass and the EoS. For EoSs with a sufficiently high $M_{\mathrm{max}}$ stable or very long-lived rigidly rotating NSs are the final product.

A prompt collapse occurs for three EoSs of our sample (marked by x in Tab.~\ref{tab:models} and~Fig.~\ref{fig:tov}). One observes this scenario only for EoSs with small $R_{\mathrm{max}}$. In the simulations with the remaining EoSs DROs are formed. The evolution of these mergers is qualitatively similar. The dynamics are described in \cite{2007A&A...467..395O,PhysRevD.81.024012}.

\begin{figure}
\includegraphics[width=8.9cm]{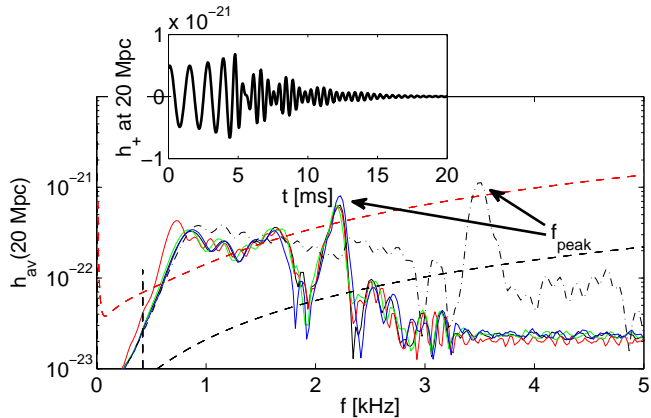}
\caption{\label{fig:ampspec}Orientation-averaged spectra of the GW signal
  for the Shen (solid) and the eosUU (black dashed-dotted) EoSs and the Adv LIGO (red dashed) and ET (black dashed) unity SNR sensitivities. The inset shows the GW amplitude with + polarization at a polar distance of 20~Mpc for the Shen EoS.}
\end{figure}

For all models that produce a DRO the GW signal is analyzed by a post-Newtonian quadrupole formula \cite{2007A&A...467..395O}. The inset of Fig.~\ref{fig:ampspec} shows the GW amplitude of the plus polarization at a polar distance of $20$~Mpc for NSs described by the Shen EoS. Clearly visible is the inspiral phase with an increasing amplitude and frequency (until~5~ms), followed by the merging and the ringdown of the postmerger remnant (from~6~ms). All DROs are stable against collapse well beyond the complete damping of the postmerger oscillations. In Fig.~\ref{fig:ampspec} we plot the spectra of the angle-averaged effective amplitude, $h_{\mathrm{av}}=0.4 f \tilde{h}_{z}(f)$ (see e.g.~\cite{2011PhRvD..83l4008H}), at a distance of 20~Mpc for the Shen EoS (solid black) and the eosUU (dash-dotted) together with the anticipated sensitivity for Adv. LIGO~\cite{2010CQGra..27h4006H} and the planned Einstein Telescope (ET)~\cite{2010CQGra..27a5003H}. Here $\tilde{h}_{z}(f)=\sqrt{(|\tilde{h}_{+}|^2+|\tilde{h}_{\times}|^2)/2}$ is given by the Fourier transforms, $\tilde{h}_{+/ \times}$, of the waveforms for both polarizations observed along the pole. As a characteristic feature of the spectra a pronounced peak at $f_{\mathrm{peak}}=2.19$~kHz for the Shen EoS and 3.50~kHz for eosUU is found, which is known to be connected to the GW emission of the merger remnant \cite{1994PhRvD..50.6247Z}. Recently, this peak has been identified as the frequency of the fundamental quadrupolar fluid mode (f-mode) \cite{2011arXiv1105.0368S}. For all models producing a DRO the spectra are sharply peaked in the kHz range around $f_{\mathrm{peak}}$ with a FWHM below 250~Hz. Values of $f_{\mathrm{peak}}$, the FWHM and the height of the peak for all models are listed in Tab.~\ref{tab:models}. For the Shen EoS Fig.~\ref{fig:ampspec} also shows the results of a run starting 3.5 revolutions before merging (red line), for a calculation with 1,270,000 SPH particles (blue), and for a simulation neglecting the GW backreaction in the postmerger phase (green) confirming the insensitivity to these choices. The initial rotation state of the NSs is known to affect $f_{\mathrm{peak}}$ only insignificantly~\cite{2007PhRvL..99l1102O}. Furthermore, our $f_{\mathrm{peak}}$ values agree within a few per cent with the results of fully relativistic simulations (e.g. 3.35~kHz for the APR EoS in \cite{2006PhRvD..73f4027S}). The uncertainties associated with the $\Gamma_{\mathrm{th}}-$ansatz for thermal effects are below 10~per cent \cite{2010PhRvD..82h4043B}.

\begin{figure}
\includegraphics[width=8.9cm]{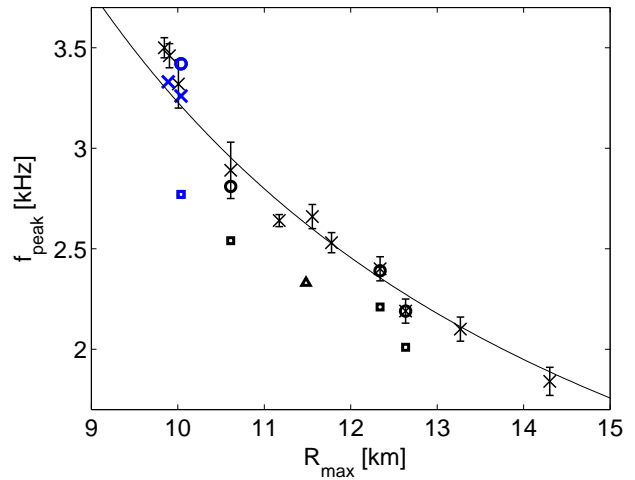}
\caption{\label{fig:frmax}Peak frequency of the postmerger GW emission vs. radius of the maximum-mass TOV solution. Blue cases are excluded by~\cite{2010Natur.467.1081D}. Error estimates are based on a Fisher matrix analysis for a source at 20~Mpc. The line is a least square fit $R_{\mathrm{max}}\propto f_{\mathrm{peak}}^{-2/3}$ for the accepted models. The triangle is an outlier (see text) and the squares correspond to models with lower $M_{\mathrm{tot}}$. See text for other symbols.}
\end{figure}

Our systematic study reveals that the peak frequency $f_{\mathrm{peak}}$ of the GW signal produced by the oscillating, hot, highly deformed DRO is determined by characteristic properties of NSs on the $M$-$R$-sequence for nonrotating TOV solutions. In Fig.~\ref{fig:frmax} $f_{\mathrm{peak}}$ is plotted against $R_{\mathrm{max}}$ (crosses and triangles) and an obvious empirical correlation is visible. $f_{\mathrm{peak}}$ is higher for smaller $R_{\mathrm{max}}$. The outlier (triangle) belongs to the simulation for the Glendnh3 EoS, which has a strikingly different $M$-$R$ relation (dashed line in Fig.~\ref{fig:tov}), which seems in conflict with theoretical knowledge of EoS properties at subnuclear densities~\cite{2010PhRvL.105p1102H}. Ignoring the outlier, the remaining ``accepted models'' exhibit an even stronger $f_{\mathrm{peak}}$-$R_{\mathrm{max}}$ correlation (line in Fig.~\ref{fig:frmax}). Already one determination of $f_{\mathrm{peak}}$ could therefore seriously constrain the $M$-$R$ relation and consequently the nuclear EoS. Additionally, simulated mergers of 1.2~$M_{\odot}$-1.5~$M_{\odot}$ binaries for selected EoSs (circles) demonstrate that the relation between $f_{\mathrm{peak}}$ and $R_{\mathrm{max}}$ is not very sensitive to the initial mass ratio~\cite{2007PhRvL..99l1102O}. Squares in Fig.~\ref{fig:frmax} display results for 1.2~$M_{\odot}$-1.2~$M_{\odot}$ mergers. For those $f_{\mathrm{peak}}$ is clearly lower~\cite{2007PhRvL..99l1102O} with differences being larger for smaller $R_{\mathrm{max}}$. But also for the symmetric binaries with lower mass a correlation seems to exist. We stress that the total binary mass $M_{\mathrm{tot}}$ is measurable by the GW inspiral signal~\cite{1994PhRvD..49.2658C}.

\begin{figure}
\includegraphics[width=8.9cm]{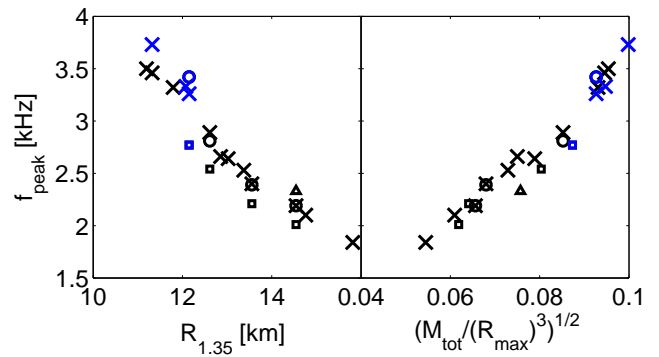}
\caption{\label{fig:fr135}Peak frequency vs. radius of a 1.35~$M_{\odot}$ NS (left) and vs. $\sqrt{M_{\mathrm{tot}}/R_{\mathrm{max}}^3}$ in geometrical units (right) with $M_{\mathrm{tot}}$ being the binary mass. Symbols have same meaning as in Fig.~\ref{fig:frmax}.}
\end{figure}

$f_{\mathrm{peak}}$ turns out to correlate also with other properties of static, cold NSs: From Fig.~\ref{fig:fr135} (left panel) a close relation between the radius $R_{1.35}$ of a 1.35~$M_{\odot}$ star and $f_{\mathrm{peak}}$ is evident. A similar coupling is found between $f_{\mathrm{peak}}$ and the maximum central density $\rho_{\mathrm{max}}$ of nonrotating NSs, where higher $\rho_{\mathrm{max}}$ yield higher $f_{\mathrm{peak}}$. However, no clear correlation exists between $f_{\mathrm{peak}}$ and $M_{\mathrm{max}}$, though typically a lower $M_{\mathrm{max}}$ gives a higher $f_{\mathrm{peak}}$, and $f_{\mathrm{peak}}>2.8$~kHz seems incompatible with $M_{\mathrm{max}}>2.4~M_{\odot}$.

A tight relation between the frequency of nonlinear oscillations of the hot, highly deformed DRO and the properties of cold, static, spherical NSs (on the $M$-$R$-sequence) is an empirical finding of this work. This links to numerical calculations which showed that for any EoS the frequency of the f-mode (generating the GW radiation at $f_{\mathrm{peak}}$~\cite{2011arXiv1105.0368S}) depends nearly linearly on the square root of the mean density, $(M/R^3)^{1/2}$, reflecting the dynamical time scale of a stellar object~\cite{1998MNRAS.299.1059A}. While $M$ here is identified with the mass of the DRO, approximately given by $M_{\mathrm{tot}}$, the empirical correlation of Fig.~\ref{fig:frmax}, however, means that $R$ in this formula can be expressed by $R_{\mathrm{max}}$ of the maximum-mass TOV configuration. With $M_{\mathrm{tot}}$ being fixed, this means that $f_{\mathrm{peak}} \propto R_{\mathrm{max}}^{-1.5}$, which is verified by the right panel of Fig.~\ref{fig:fr135}, where except for the mentioned outlier a clear linear scaling is visible. A fit to $R_{\mathrm{max}}(f_{\mathrm{peak}}) \propto f_{\mathrm{peak}}^{-2/3}$ (line in Fig.~\ref{fig:frmax}), using only accepted models, reveals a maximum residual of 300~m. The tight correlation of $f_{\mathrm{peak}}$ and $(M_{\mathrm{tot}}/R_{\mathrm{max}}^3)^{1/2}$ implies that the radius of the DRO and $R_{\mathrm{max}}$ are strongly linked. Such a strong link has indeed been empirically found to exist between $R_{\mathrm{max}}$ and the radius of the most massive, rigidly rotating NS for any EoS~\cite{1996ApJ...456..300L}, and seems to exist also for differentially rotating NSs with 2.7~$M_{\odot}$.

Using the postmerger signal alone and correcting the underestimation of 40~\% of the GW amplitude by the quadrupole formula~\cite{2005PhRvD..71h4021S}, a SNR of 2 (the inclusion of the inspiral signal increases the SNR significantly) yields an optimal detection horizon of about 20--45~Mpc (dependent on the EoS) for Adv. LIGO. This corresponds to 145--1190 Milky Way Equivalent Galaxies accessible for a GW search, taking into account the reduction due to random source location and orientation~\cite{2010CQGra..27q3001A}. The ``realistic'' and the ``high'' merger rates of~\cite{2010CQGra..27q3001A} imply a detection rate of 0.01--1~events/yr for Adv. LIGO. With the planned ET \cite{2010CQGra..27a5003H} and its higher sensitivity several observations of $f_{\mathrm{peak}}$ per year will become very likely.

For polar distances of 20~Mpc $f_{\mathrm{peak}}$-uncertainties of typically 50~Hz and at most 160~Hz can be estimated from the Fisher information matrix for neighboring cases of accepted models following~\cite{2009PhRvD..79l4033R}. Corresponding uncertainties are indicated in Fig.~\ref{fig:frmax} by averages for contiguous pairs of models. Considering in addition the residuals to the fits of the relations of Figs.~\ref{fig:frmax} and~\ref{fig:fr135}, a measurement of $f_{\mathrm{peak}}$ will allow to determine the NS radius with an accuracy of several 100~m. These prospects are comparable with the 1~km accuracy of the radius estimation for the initial NSs from the inspiral GW signal of symmetric binaries suggested in~\cite{2009PhRvD..79l4033R} for events within a maximal distance of 20--100~Mpc. Both will set strong constraints on the $M$-$R$ relation and thus the EoS~\cite{2007PhR...442..109L}. Our approach, however, is an independent, complementary one, also to the possibility of gaining EoS information from the weak correlation between $M_{\mathrm{max}}$ and the threshold total binary mass $M_{\mathrm{thres}}$ that distinguishes prompt $(M_{\mathrm{tot}}>M_{\mathrm{thres}})$ from delayed $(M_{\mathrm{tot}}<M_{\mathrm{thres}})$ BH formation~\cite{2011PhRvD..83l4008H}, whose determination requires more than one GW detection~\cite{2005PhRvL..94t1101S}.

Future numerical studies should vary $M_{\mathrm{tot}}$ and confirm our findings by more sophisticated models of binary mergers, e.g. considering magnetic fields, neutrino physics, and full general relativity. Also the detectability of $f_{\mathrm{peak}}$ should be explored in more detail, e.g. by a detector network. Finally, our explanation should be examined more closely to develop a precise understanding of the presented correlations.

\begin{acknowledgments}
We thank N.~Stergioulas, D.~Shoemaker and S.~Hild. This work was supported by DFG grants SFB/TR~7, SFB/TR~27, EXC~153, by ESF/CompStar, and by computer time at LRZ Munich, RZG Garching and MPA.
\end{acknowledgments}

\bibliography{references}

\end{document}